\documentclass[preprint,showpacs,superscriptaddress,nofootinbib,prd]{revtex4}

\usepackage{latexsym}
\usepackage{amssymb}
\usepackage[dvips]{graphicx}

\newcommand{\tr}{\mathop{\textrm{tr}}}

\begin{document}  

\preprint{\parbox[t]{15em}{\raggedleft
FERMILAB-PUB-02/178-T \\
KEK-CP-127 \\
HUPD-0204 \\
YITP-02-47 \\
hep-lat/0208004}}

\title{\boldmath\vspace*{1.0in}
Perturbative calculation of $O(a)$ improvement coefficients}

\author{Junpei Harada}
\affiliation{Department of Physics, Hiroshima University,
	Higashi-Hiroshima 739-8526, Japan}
\author{Shoji Hashimoto}
\affiliation{High Energy Accelerator Research Organization (KEK),
	Tsukuba 305-0801, Japan}
\author{Andreas S. Kronfeld}
\affiliation{Theoretical Physics Department, Fermi National
	Accelerator Laboratory, \\ Batavia, Illinois 60510, USA}
\author{Tetsuya Onogi}
\affiliation{Yukawa Institute for Theoretical Physics,
	Kyoto University, \\
	Sakyo-ku, Kyoto 606-8502, Japan}

\date{August 1, 2002}

\begin{abstract}
We compute several coefficients needed for $O(a)$ improvement of
currents in perturbation theory, using the Brodsky-Lepage-Mackenzie
prescription for choosing an optimal scale~$q^*$.
We then compare the results to non-perturbative calculations.
Normalization factors of the vector and axial vector currents show good
agreement, especially when allowing for small two-loop effects.
On the other hand, there are large discrepancies in the coefficients of
$O(a)$ improvement terms.
We suspect that they arise primarily from power corrections inherent
in the non-perturbative methods.
\end{abstract} 


\maketitle


\section{Introduction}

During the past few years the Symanzik effective field theory has been
an important focus of research in lattice gauge theory.
Symanzik's idea is to describe cutoff effects in lattice field theory
by a continuum effective field theory~\cite{Symanzik:1979ph}.
One writes~\cite{Symanzik:1979ph,Symanzik:1983dc}
\begin{equation}
	{\cal L}_{\text{lat}} \doteq {\cal L}_{\text{Sym}},
\end{equation}
where the symbol $\doteq$ means that the lattice and Symanzik field
theories have the same on-shell matrix elements.
For lattice QCD with Wilson fermions~\cite{Wilson:1975hf} the
Symanzik local effective Lagrangian (LE${\cal L}$) is given
by~\cite{Sheikholeslami:1985ij,Kronfeld:2002pi}
\begin{equation}
	 {\cal L}_{\text{Sym}} =
		\frac{1}{2g^2}\tr[F^{\mu\nu}F_{\mu\nu}] -
		\bar{q}\left({\kern+0.1em /\kern-0.65em D} + m\right)q
		 + aK_{\sigma\cdot F}\, 
		\bar{q}i\sigma_{\mu\nu}F^{\mu\nu}q + O(a^2), 
	\label{eq:LEL}
\end{equation}
where $g^2$ is a renormalized coupling,
$m$ is a renormalized quark mass,
and $aK_{\sigma\cdot F}$ is a short-distance coefficient.
The effective field theory is useful when the scale of QCD in lattice
units is small, $\Lambda a\ll1$, and,
as used in this paper, when $ma\ll1$~also.
With the description in hand, the lattice field theory can be adjusted
so that it approaches its continuum limit more quickly.
The effective theory shows that if $K_{\sigma\cdot F}$ is
reduced for any given on-shell matrix element, then the $O(a)$ term in
Eq.~(\ref{eq:LEL}) makes commensurately smaller contributions to all
other on-shell matrix elements.
This application of the Symanzik effective field theory is called the
Symanzik improvement program~\cite{Symanzik:1983dc}.

A similar correspondence is set up for the vector and axial vector
currents (see below), introducing further short-distance coefficients.
In the last several years methods have been devised to study all of
them non-perturbatively \cite{Jansen:1996ck,Luscher:1996sc,%
Martinelli:1997zc,deDivitiis:1997ka,Bhattacharya:1999uq}.
The $O(a)$ discretization effects violate chiral symmetry, so the
key idea is to ensure that violations of chiral symmetry are at
least~$O(a^2)$.
On the other hand, because of asymptotic freedom and the success
of perturbative QCD, even at GeV energies~\cite{Davier:1997kw},
one expects perturbation theory to yield accurate estimates of the
short-distance coefficients.
In this paper, we compare a perturbative calculation of the currents'
short-distant coefficients to the non-perturbative results.

There are two issues that should be kept in mind when making such a
comparison.
First, the non-perturbative technique suffers from power corrections.
Asymptotically, as $\Lambda a\to0$ these are formally smaller than any
error made from truncating the perturbation series.
In practice, however, these effects can be significant.

Second, no two-loop results are available for the improvement
coefficients considered here.
Tests of perturbation theory are, therefore, not unambiguous, because
different choices for the expansion parameter~$g^2$ yield quantitatively
different results.
The bare coupling~$g_0^2$ (for the Wilson gauge action) is an especially
bad expansion parameter~\cite{Lepage1991:zt}.
The obvious remedy is to rearrange the perturbative series,
eliminating $g_0^2$ in favor of a renormalized (running) coupling,
evaluated at a scale characteristic of the problem at hand.
One is then faced, however, with many choices of renormalization scheme,
and the question of how to determine the ``characteristic scale.''
In this paper we choose the Brodsky-Lepage-Mackenzie (BLM)
prescription~\cite{Brodsky:1983gc,Lepage:1993xa}.
Once this choice is made, little subjectivity remains, so one can ask
quantitatively whether one-loop BLM perturbation theory agrees with
the non-perturbative method.

In the BLM method, the characteristic scale is computed from Feynman
diagrams.
The new information presented in this paper consists of the calculations
needed to determine the BLM scales of the normalization and improvement
coefficients of the vector and axial vector currents for Wilson fermions
with Sheikholeslami-Wohlert action.
These calculations are a by-product of our recent work on the
normalization and improvement of lattice currents with heavy
quarks~\cite{Harada:2001fi}.
Details of the calculational method may be found there.

This paper is organized as follows.
In Sec.~\ref{sec:currents} we define the lattice currents 
and review their description in the Symanzik effective field theory.
Section~\ref{sec:BLM} recalls the BLM prescription, focusing on points
that are sometimes overlooked.
Our new results for the BLM scales are given in Sec.~\ref{sec:results}.
This paves the way for a systematic comparison with non-perturbative
calculations of the same quantities in Sec.~\ref{sec:comparison}.
Section~\ref{sec:conclusions} contains a few concluding remarks.

\section{Lattice Currents}
\label{sec:currents}

In this section we review the description of lattice currents with the
Symanzik effective field theory.
For quarks we take the Sheikholeslami-Wohlert
Lagrangian~\cite{Sheikholeslami:1985ij},
which has an improvement coupling~$c_{\text{SW}}$.
At the tree level
$K_{\sigma\cdot F}^{[0]}=\frac{1}{4}(1-c_{\text{SW}})$,
so the improvement condition $K_{\sigma\cdot F}=0$
requires $c_{\text{SW}}=1+O(g^2)$.
For one-loop calculations, it is sufficient to specify $c_{\text{SW}}$
at the tree level.
For the non-perturbative calculations cited below,
$c_{\text{SW}}-1$ is determined non-perturbatively
by the methods of Ref.~\cite{Luscher:1996sc}.

We denote the lattice fermion field with~$\psi$.
The lattice vector and axial vector currents take the form
\begin{eqnarray}
	V^\mu_{\text{lat}} & = & \bar{\psi}i\gamma^\mu\psi -
		ac_V{\partial_\nu}_{\text{lat}}\bar{\psi}\sigma^{\mu\nu}\psi,
	\label{eq:Vlat} \\
	A^\mu_{\text{lat}} & = & \bar{\psi}i\gamma^\mu\gamma_5\psi +
		ac_A \partial^\mu_{\text{lat}} \bar{\psi}i\gamma_5\psi.
	\label{eq:Alat}
\end{eqnarray}
The improvement couplings $c_V$ and $c_A$ should be chosen to reduce
lattice artifacts, as discussed below.%
\footnote{The lattice currents in Eqs.~(\ref{eq:Vlat})
and~(\ref{eq:Alat}) are useful for light quarks.
For heavy quarks the ``small'' improvement terms become large,
introducing unnecessary violations of heavy-quark symmetry.
Better currents for heavy quarks are given in
Refs.~\cite{Harada:2001fi,Harada:2001fj}.}
In Symanzik's theory of cutoff effects, the lattice currents
are described by operators in a continuum effective field theory~%
\cite{Symanzik:1979ph,Symanzik:1983dc,Kronfeld:2002pi,Luscher:1996sc}
\begin{eqnarray}
	V^\mu_{\text{lat}} & \doteq &
		\bar{Z}_V^{-1}    \bar{q}i\gamma^\mu         q -
		aK_V \partial_\nu \bar{q} \sigma^{\mu\nu}    q + \cdots ,
	\label{eq:LEV} \\
	A^\mu_{\text{lat}} & \doteq &
		\bar{Z}_A^{-1}    \bar{q}i\gamma^\mu\gamma_5 q +
		aK_A \partial^\mu \bar{q}i          \gamma_5 q + \cdots ,
	\label{eq:LEA}
\end{eqnarray}
where, as in Eq.~(\ref{eq:LEL}), $q$ is a continuum fermion field
whose dynamics is defined by ${\cal L}_{\text{QCD}}$.
The ellipsis indicates operators of dimension five and higher.
Further dimension-four operators are omitted from 
Eqs.~(\ref{eq:LEV}) and~(\ref{eq:LEA}), because they are linear 
combinations of those listed and others that vanish by the equations 
of motion.
The short-distance coefficients in the effective Lagrangian---%
$\bar{Z}_J$ and $K_J$ ($J=V$, $A$)---are functions of $g^2$ and $ma$,
and the improvement couplings $c_{\text{SW}}$ and $c_J$.

Symanzik improvement is achieved by adjusting $c_J$ so that $K_J=0$.
Then $\bar{Z}_VV^\mu_{\text{lat}}$ and $\bar{Z}_AA^\mu_{\text{lat}}$
have the same matrix elements as $\bar{q}i\gamma^\mu q$
and $\bar{q}i\gamma^\mu\gamma_5q$, apart from lattice artifacts of
order~$a^2$.
For light quarks one may expand $\bar{Z}_J$ in $ma$,
\begin{equation}
	\bar{Z}_J = Z_J\left(1 + mab_J\right),
\end{equation}
and identify $K_J$ with the zeroth order of a small $ma$ expansion.
At the tree level the coefficients of the normalization factor are
$Z_J^{[0]}=1$, $b_J^{[0]}=1$.
In addition, the coefficient of the lattice artifact is
\begin{equation}
	K_J^{[0]} = c_J^{[0]}.
\end{equation}
The improvement condition $K_J=0$ says that one should set
$c_J^{[0]}=0$.
Consequently, one-loop calculations are based solely on the first
terms in Eqs.~(\ref{eq:Vlat}) and~(\ref{eq:Alat}).

\section{Brodsky-Lepage-Mackenzie Prescription}
\label{sec:BLM}

In this section we review the BLM prescription,
following the argumentation from Ref.~\cite{Lepage:1993xa}.
This material should be familiar, but some of the literature on
non-perturbative improvement blurs the difference between BLM
perturbation theory and other topics, such as ``tadpole improvement''
and mean-field estimates of the renormalized coupling,
which are also discussed in Ref.~\cite{Lepage:1993xa}.

The problem is to find a reasonably accurate one-loop estimate of
a quantity~$\zeta$, here $\bar{Z}_J$ or $K_J$.
In these cases, one gluon with momentum~$k$ and propagator~$D(k)$
appears.
The contribution from the Feynman diagrams can be written
\begin{equation}
	g^2_R \zeta^{[1]}(p) =
		g_0^2 \int \frac{d^4k}{(2\pi)^4} D(k) f(k,p) + \cdots,
\end{equation}
where $p$ denotes $k$-independent parameters, such as external momenta.
The ellipsis indicates higher-order terms that we would like to absorb
into the renormalized coupling~$g^2_R$.
An important class of higher-order terms consists of the renormalization
parts that dress the exchanged gluon.
In the Fourier transform of the heavy-quark potential, for example,
they turn $g_0^2D(k)$ into $g_V^2(k)D(k)$, where the potential
$V(q)=-C_Fg^2_V(q)/q^2$.
Thus,
\begin{equation}
	g^2_R \zeta^{[1]}(p) =
		\int \frac{d^4k}{(2\pi)^4} g_V^2(k) D(k) f(k,p) + \cdots
	\label{eq:one-loop-renormed}
\end{equation}
sums the renormalization parts.
Other ways of dressing the gluon would lead to other physical
running couplings, but they all are the same at order~$\beta_0g^4$
\cite{Brodsky:1983gc}, where $\beta_0=11-\frac{2}{3}n_f$ is the one-loop
coefficient of the $\beta$~function for $n_f$ light quarks.

If there is a characteristic scale $q^*$, one can approximate
\begin{eqnarray}
	g_V^2(k) & = &
		\frac{g_V^2(q^*)}{1+(\beta_0/16\pi^2)g_V^2(q^*)\ln(k/q^*)^2} \\
		& = & g_V^2(q^*) +
		\frac{\beta_0}{16\pi^2}g_V^4(q^*)\ln(q^*/k)^2 + \cdots.
	\label{eq:gVexpanded}
\end{eqnarray}
The aim is to choose $q^*$ so that higher-order terms are small,
particularly those of order $\beta_0g_V^4$, which could be
enhanced by a foolish choice of $q^*$.
Inserting Eq.~(\ref{eq:gVexpanded}) into
Eq.~(\ref{eq:one-loop-renormed}) and setting the coefficient of
$\beta_0g_V^4$ to zero yields
\begin{equation}
	\ln q^*a = \frac{{}^*\zeta^{[1]}}{2\zeta^{[1]}},
	\label{eq:qstarV}
\end{equation}
where $a$ is a reference short-distance scale
(namely, the lattice spacing), and
\begin{equation}
	{}^*\zeta^{[1]}(p) =
		\int \frac{d^4k}{(2\pi)^4} \ln(ka)^2 D(k) f(k,p).
\end{equation}
Thus, the BLM prescription is to set $g_R^2=g_V^2(q^*)$ in
the one-loop approximation.

If one prefers a different renormalized coupling,
one must change the scale in the appropriate way.
The coupling in scheme ``$S$'' is related to the $V$~scheme by
\begin{equation}
	\frac{1}{g^2_S(q)} = \frac{1}{g^2_V(q)} +
		\frac{\beta_0 b_S^{(1)} + b_S^{(0)}}{16\pi^2} + O(g^2),
	\label{eq:bS}
\end{equation}
where $b_S^{(0)}$ and $b_S^{(1)}$ are constants independent of~$n_f$.
The BLM scale~$q^*_S$ for this scheme is given by
\begin{equation}
	\ln q^*_S = \ln q^* - \frac{1}{2}b_S^{(1)}.
	\label{eq:qstarS}
\end{equation}
For example, for the modified minimal subtraction ($\overline{\rm MS}$)
scheme, $b_{\overline{\rm MS}}^{(0)}=-8$ and
$b_{\overline{\rm MS}}^{(1)}=5/3$,
$q^*_{\overline{\rm MS}}=e^{-5/6}q^*=0.435q^*$.
With Eq.~(\ref{eq:qstarS}) one recovers the summary statement of
Ref.~\cite{Brodsky:1983gc}, namely to absorb into $q^*_S$ the $n_f$
dependence of the two-loop term, which enters through~$\beta_0$.

The BLM prescription has several features that make it a natural choice
in matching calculations, such as those considered in this paper.
The effective field theory framework suggests using a renormalized
coupling, in particular one that has a (quasi-)physical definition in
both the underlying theory (here lattice gauge theory) and in the
effective theory (here the Symanzik effective field theory).
For quantitative purposes it is more interesting to note that
\begin{equation}
	\frac{1}{g^2_S(q^*_S)} = \frac{1}{g^2_V(q^*)} +
		\frac{b_S^{(0)}}{16\pi^2} + O(g^2),
	\label{eq:MS}
\end{equation}
so the numerical difference in the BLM expansion parameters is small, as
long as $g^2b_S^{(0)}/16\pi^2$ is small.

\section{Perturbative Results}
\label{sec:results}

In Ref.~\cite{Harada:2001fi} we found for gauge group SU(3)
and $c_{\text{SW}}=1$
\begin{eqnarray}
    Z_V^{[1]} & = & - 0.129423(6), \label{eq:ZV1} \\
    Z_A^{[1]} & = & - 0.116450(5), \label{eq:ZA1}
\end{eqnarray}
in excellent agreement with previous
work~\cite{Gabrielli:1990us,Capitani:2001xi}.
(Reference~\cite{Capitani:2001xi} gives precise results as
a polynomial in~$c_{\text{SW}}$.) 
We also found (with $c_J^{[0]}=0$)
\begin{eqnarray}
    b_V^{[1]}    & = & 0.153239(14), \label{eq:bV1} \\
    b_A^{[1]}    & = & 0.152189(14), \label{eq:bA1} \\
    K_V^{[1]}    & = & c_V^{[1]} + 0.016332(7),   \label{eq:KV1} \\
    K_A^{[1]}    & = & c_A^{[1]} + 0.0075741(15), \label{eq:KA1}
\end{eqnarray}
which agree perfectly with Ref.~\cite{Sint:1997jx}.
Solving the improvement condition $K_J=0$ at this order gives
\begin{eqnarray}
    c_V^{[1]}    & = & - 0.016332(7),   \label{eq:cV1} \\
    c_A^{[1]}    & = & - 0.0075741(15). \label{eq:cA1}
\end{eqnarray}
We also directly obtained
\begin{equation}
    b_V^{[1]} - b_A^{[1]} = 0.0010444(16),
    \label{eq:bV-bA}
\end{equation}
which is more accurate than the difference of the two numbers
quoted above.
In taking the difference, large contributions from the self energy
cancel, but, even so, the near equality of~$b_V^{[1]}$ and~$b_A^{[1]}$
is a bit astonishing.
The mass dependence of $\bar{Z}_J$ shows that $b_V^{[1]}-b_A^{[1]}$
is not so small for the Wilson action~\cite{Harada:2001fi}.

In our method for computing the improvement coefficients it is easy to
weight the integrands with $\ln(ka)^2$ and, thus, obtain the BLM scales.
We find
\begin{eqnarray}
	{}^*Z_V^{[1]} & = & - 0.270691(19), \label{eq:starZV} \\
	{}^*Z_A^{[1]} & = & - 0.243086(09), \\
	{}^*b_V^{[1]} & = &   0.321556(35), \\
	{}^*b_A^{[1]} & = &   0.318108(21), \\
	{}^*b_V^{[1]} - {}^*b_A^{[1]} & = & 0.0034247(51), \\
	{}^*c_V^{[1]} & = & - 0.0222383(15), \\
	{}^*c_A^{[1]} & = & - 0.0147825(62), \label{eq:starcA}
\end{eqnarray}
and hence
\begin{eqnarray}
	q^*_{Z_V}a & = & 2.846, \\
	q^*_{Z_A}a & = & 2.840, \\
	q^*_{Z_A/Z_V}a & = & 2.898, \\
	q^*_{b_V}a & = & 2.855, \\
	q^*_{b_A}a & = & 2.844, \\
	q^*_{b_V-b_A}a & = & 5.153, \\
	q^*_{c_V}a & = & 1.975, \\
	q^*_{c_A}a & = & 2.653.
\end{eqnarray}
The scales are in the expected range.
The higher scale for $b_V-b_A$ means simply that the difference between
these renormalization constants arises from very short distances.
These numerical results are new; they have been obtained from two
independent computer programs.
As a further check, we have reproduced the values of $q^*_{Z_V}a$
and $q^*_{Z_A}a$ for the Wilson action ($c_{\text{SW}}=0$),
given in Ref.~\cite{Bernard:1998sx}.

The dominant contributor to the ``large'' one-loop normalization
constants, Eqs.~(\ref{eq:ZV1})--(\ref{eq:bA1}), is the tadpole diagram
(in Feynman gauge) of the self energy.
One might expect perturbation theory to work better for quantities in
which the effects of tadpole diagrams largely cancel (albeit in a
gauge-invariant way).
For example, $Z_A/Z_V$ and $b_V-b_A$ are tadpole free and have smaller
one-loop coefficients.

Another way to remove the tadpoles is to write
\begin{eqnarray}
	Z_J = u_0 \tilde{Z}_J, \\
	b_J = \tilde{b}_J/u_0,
\end{eqnarray}
where $u_0$ is any convenient tadpole-dominated quantity.
Then one can take $u_0$ from a non-perturbative Monte Carlo calculation
and use perturbation theory for $\tilde{Z}_J$ and~$\tilde{b}_J$.
The corresponding one-loop coefficients are
\begin{eqnarray}
	\tilde{Z}_J^{[1]} = Z_J^{[1]} - u_0^{[1]}, \label{eq:ZJtad} \\
	\tilde{b}_J^{[1]} = b_J^{[1]} + u_0^{[1]}.
\end{eqnarray}
Similarly, to get the BLM scale
\begin{eqnarray}
	{}^*\tilde{Z}_J^{[1]} = {}^*Z_J^{[1]} - {}^*u_0^{[1]}, \\
	{}^*\tilde{b}_J^{[1]} = {}^*b_J^{[1]} + {}^*u_0^{[1]},
	\label{eq:starbJtad}
\end{eqnarray}
where $^*u_0^{[1]}$ is the BLM numerator [cf.\ Eq.~(\ref{eq:qstarV})]
for $u_0$.
Below we take $u_0^4$ to be the average value of the plaquette,
with $u_0^{[1]}=-1/12=-0.08\bar{3}$ and $^*u_0^{[1]}=-0.204049(1)$.
A~glance at Eqs.~(\ref{eq:ZJtad})--(\ref{eq:starbJtad}) shows
immediately that tadpole improvement reduces the one-loop coefficients.
With tadpole improvement the BLM scales become
\begin{eqnarray}
	q^*_{\tilde{Z}_V}a & = & 2.061, \\
	q^*_{\tilde{Z}_A}a & = & 1.803, \\
	q^*_{\tilde{b}_V}a & = & 2.317, \\
	q^*_{\tilde{b}_A}a & = & 2.289.
\end{eqnarray}
The scales are lower than without tadpole improvement, but still
ultraviolet.

It is perhaps worthwhile emphasizing the difference between
tadpole improvement and the BLM prescription.
The aim of tadpole improvement is to re-sum large contributions appearing
at order $g^2$ and higher, replacing the sum with a non-perturbative
estimate ($u_0$, for example).
The aim of the BLM prescription is to re-sum potentially large
renormalization parts into the renormalized coupling.
Although the aims are similar, they are not identical.
They are not mutually exclusive, and neither is a substitute
for the other.

\section{Comparison to Non-perturbative Calculations}
\label{sec:comparison}

\begin{table}[tbp]
	\centering
	\caption[tab:bK62]{Comparison of perturbative and non-pertur\-bative 
	determinations of the improvement coefficients at $\beta=6.2$.}
	\begin{ruledtabular}
	\begin{tabular}{cllll}
		$\beta=6.2$ & $\alpha_V(q^*)$ &
		\multicolumn{1}{c}{BLM} & 
		\multicolumn{1}{c}{Refs.~\cite{Luscher:1996ck,Luscher:1996ax,Guagnelli:1997db}} &
		\multicolumn{1}{c}{Ref.~\cite{Bhattacharya:2001pn}} \\
		\hline
		  $Z_V$   & 0.1468 & 0.7612 & 0.7922(9) & 0.7874(4) \\
		  $Z_A$   & 0.1469 & 0.7850 & 0.807(8)  & 0.818(5)  \\
		$Z_A/Z_V$ & 0.1461 & 1.0238 & 1.019(8)  & 1.039(5)  \\
		  $b_V$   & 0.1467 & 1.2824 & 1.41(2)   & 1.42(1)   \\
		  $b_A$   & 0.1468 & 1.2808 &    ---    & 1.32(5)   \\
		$b_V-b_A$ & 0.1257 & 0.001649 &  ---    & 0.11(5)   \\
		  $-c_V$  & 0.1638 & 0.03361 & 0.21(7)  & 0.09(2)   \\
		  $-c_A$  & 0.1498 & 0.01426 & 0.038(4) & 0.032(7)  \\
		\hline
$u_0 \tilde{Z}_V$ & 0.1616 & 0.8022 & 0.7922(9) & 0.7874(4) \\
$u_0 \tilde{Z}_A$ & 0.1686 & 0.8230 & 0.807(8)  & 0.818(5)  \\
$\tilde{b}_V/u_0$ & 0.1559 & 1.2846 & 1.41(2)   & 1.42(1)   \\
$\tilde{b}_A/u_0$ & 0.1565 & 1.2828 &    ---    & 1.32(5)   \\
	\end{tabular}
	\end{ruledtabular}
	\label{tab:bK62}
\end{table}
\begin{table}[tbp]
	\centering
	\caption[tab:bK60]{Comparison of perturbative and non-pertur\-bative 
	determinations of the improvement coefficients at $\beta=6.0$.}
	\begin{ruledtabular}
	\begin{tabular}{cllll}
		$\beta=6.0$ & $\alpha_V(q^*)$ &
		\multicolumn{1}{c}{BLM} & 
		\multicolumn{1}{c}{Refs.~\cite{Luscher:1996ck,Luscher:1996ax,Guagnelli:1997db}} &
		\multicolumn{1}{c}{Ref.~\cite{Bhattacharya:2001pn}} \\
		\hline
		  $Z_V$   & 0.1602 & 0.7394 & 0.7809(6) & 0.770(1)~ \\
		  $Z_A$   & 0.1603 & 0.7654 & 0.791(9)  & 0.807(8)  \\
		$Z_A/Z_V$ & 0.1593 & 1.0260 & 1.012(9)  & 1.048(8)  \\
		  $b_V$   & 0.1601 & 1.3082 & 1.54(2)   & 1.52(1)   \\
		  $b_A$   & 0.1603 & 1.3065 &    ---    & 1.28(5)   \\
		$b_V-b_A$ & 0.1352 & 0.001774 &  ---    & 0.24(5)   \\
		  $-c_V$  & 0.1808 & 0.03711 & 0.32(7)  & 0.107(17) \\
		  $-c_A$  & 0.1638 & 0.01559 & 0.083(5) & 0.037(9)  \\
		\hline
$u_0 \tilde{Z}_V$ & 0.1782 & 0.7872 & 0.7809(6) & 0.770(1)~ \\
$u_0 \tilde{Z}_A$ & 0.1868 & 0.8095 & 0.791(9)  & 0.807(8)  \\
$\tilde{b}_V/u_0$ & 0.1712 & 1.3105 & 1.54(2)   & 1.52(1)   \\
$\tilde{b}_A/u_0$ & 0.1719 & 1.3087 &    ---    & 1.28(5)   \\
	\end{tabular}
	\end{ruledtabular}
	\label{tab:bK60}
\end{table}
With the BLM scales in hand we can compare the prediction of one-loop
BLM-improved perturbation theory with non-perturbative determinations
of the improvement coefficients.
We shall make the comparison in two ways.
First we compare the numerical values directly, at two values of the
bare coupling.
Here there are two methods in the literature, one
based on finite-size techniques and the Schr\"odinger
functional~\cite{Luscher:1996ck,Luscher:1996ax,Guagnelli:1997db},
and another based on large volumes with hadronic matrix
elements~\cite{Bhattacharya:2001pn}.
The difference between these two illustrates how large power corrections
to the improvement coefficients are.
We also compare our results graphically, as a
function of coupling, to Pad\'e approximants given in
Refs.~\cite{Sint:1997jx,Luscher:1996ck,Luscher:1996ax}.
These graphs are helpful for seeing whether discrepancies in the
one-loop and non-perturbative estimates arise from two-loop or power
corrections.

We obtain $\alpha_V(q^*)$ as follows.
First we compute
\begin{equation}
	\alpha_{1\times1} = -\frac{3}{4\pi}\ln\langle\Box\rangle,
	\label{eq:1x1}
\end{equation}
where $\langle\Box\rangle$ is the ensemble average of the plaquette.
Then we follow Ref.~\cite{Lepage:1993xa} and take~$\alpha_V$ to be
\begin{equation}
	\alpha_V(3.402/a) \equiv \frac{2\alpha_{1\times1}}%
		{1+\sqrt{1-4.741\alpha_{1\times1}}},
	\label{eq:alphaVdef}
\end{equation}
which
agrees with the standard definition of $\alpha_V$ with an accuracy of
order~$\alpha_s^3$.
The scale $3.402/a$ is the BLM scale for $\langle\Box\rangle$.
We then run from~$3.402/a$ to $q^*$ with the two-loop
evolution equation.
Of course, once two-loop perturbation theory is available, one would
have to extend the accuracy of Eq.~(\ref{eq:alphaVdef}) and of the
evolution.

Table~\ref{tab:bK62} gives results from our perturbative
calculation with non-perturbative results from the {\sl Alpha}
Collaboration~\cite{Luscher:1996ck,Luscher:1996ax,Guagnelli:1997db}
and from Bhattacharya \emph{et al.}~\cite{Bhattacharya:2001pn},
at $\beta=6.2$.
Table~\ref{tab:bK60} gives the same comparison at $\beta=6.0$.
Above (below) the horizontal line, we have applied the BLM prescription
without (with) tadpole improvement.
The error bars on the entries from Refs.~\cite{Luscher:1996ck,%
Luscher:1996ax,Guagnelli:1997db,Bhattacharya:2001pn} are statistical,
and compiled in Ref.~\cite{Bhattacharya:2001pn}.

\begin{figure*}[btp]
	(a)\hspace*{-1.2em}
	\includegraphics[width=0.48\textwidth]{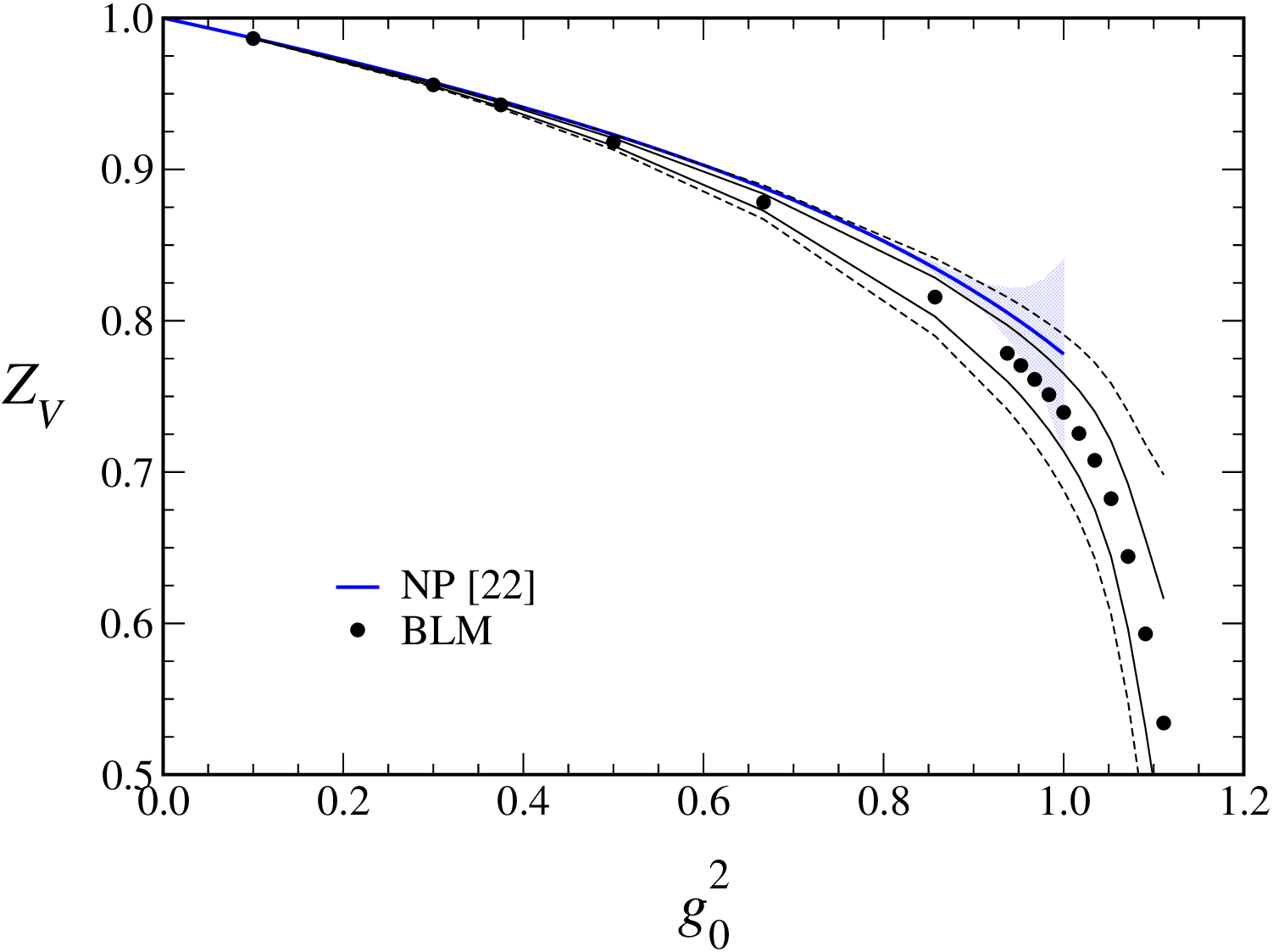} \hfill
	(b)\hspace*{-1.2em}
	\includegraphics[width=0.48\textwidth]{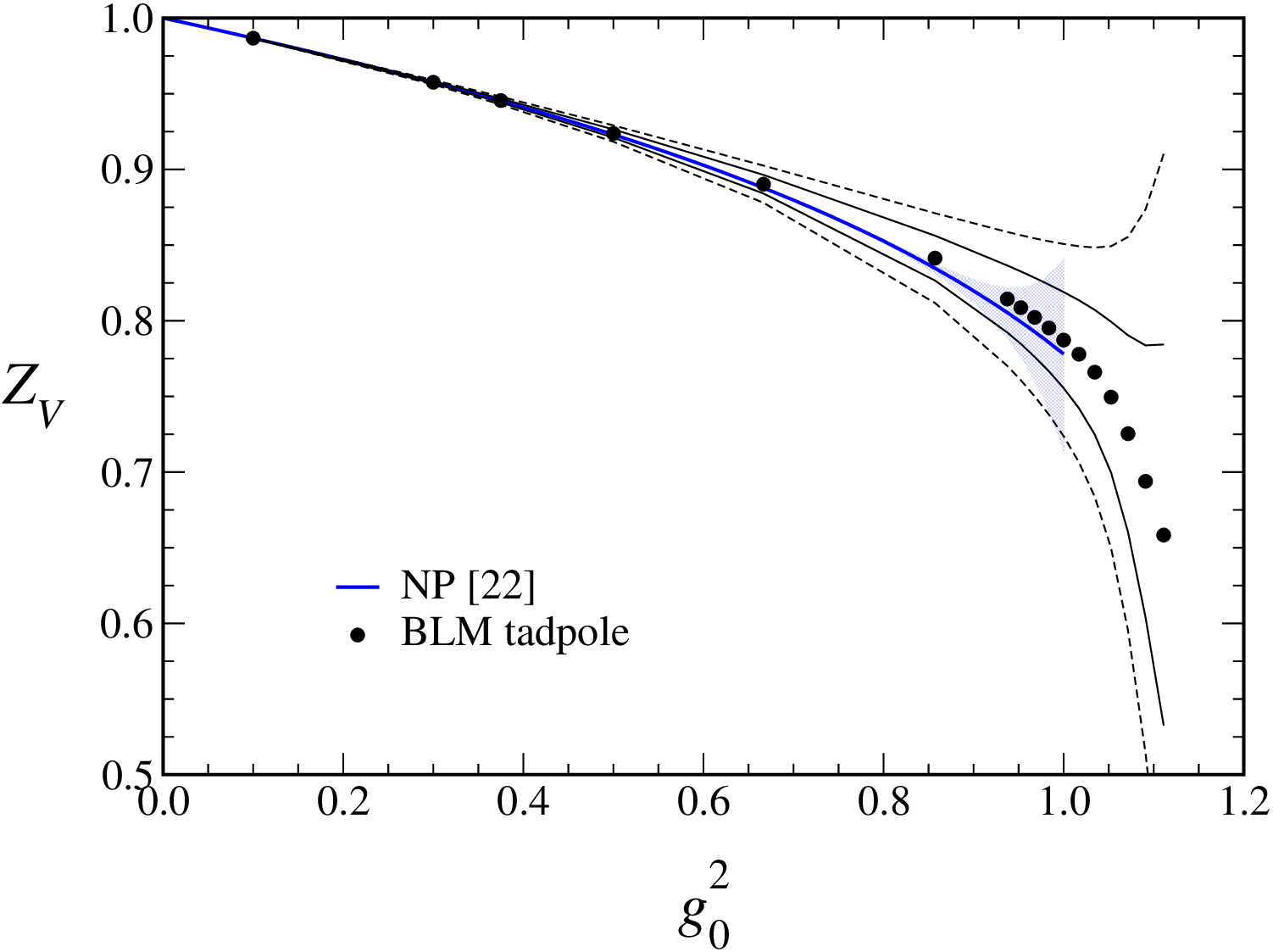}\\[0.5em]
	(c)\hspace*{-1.2em}
	\includegraphics[width=0.48\textwidth]{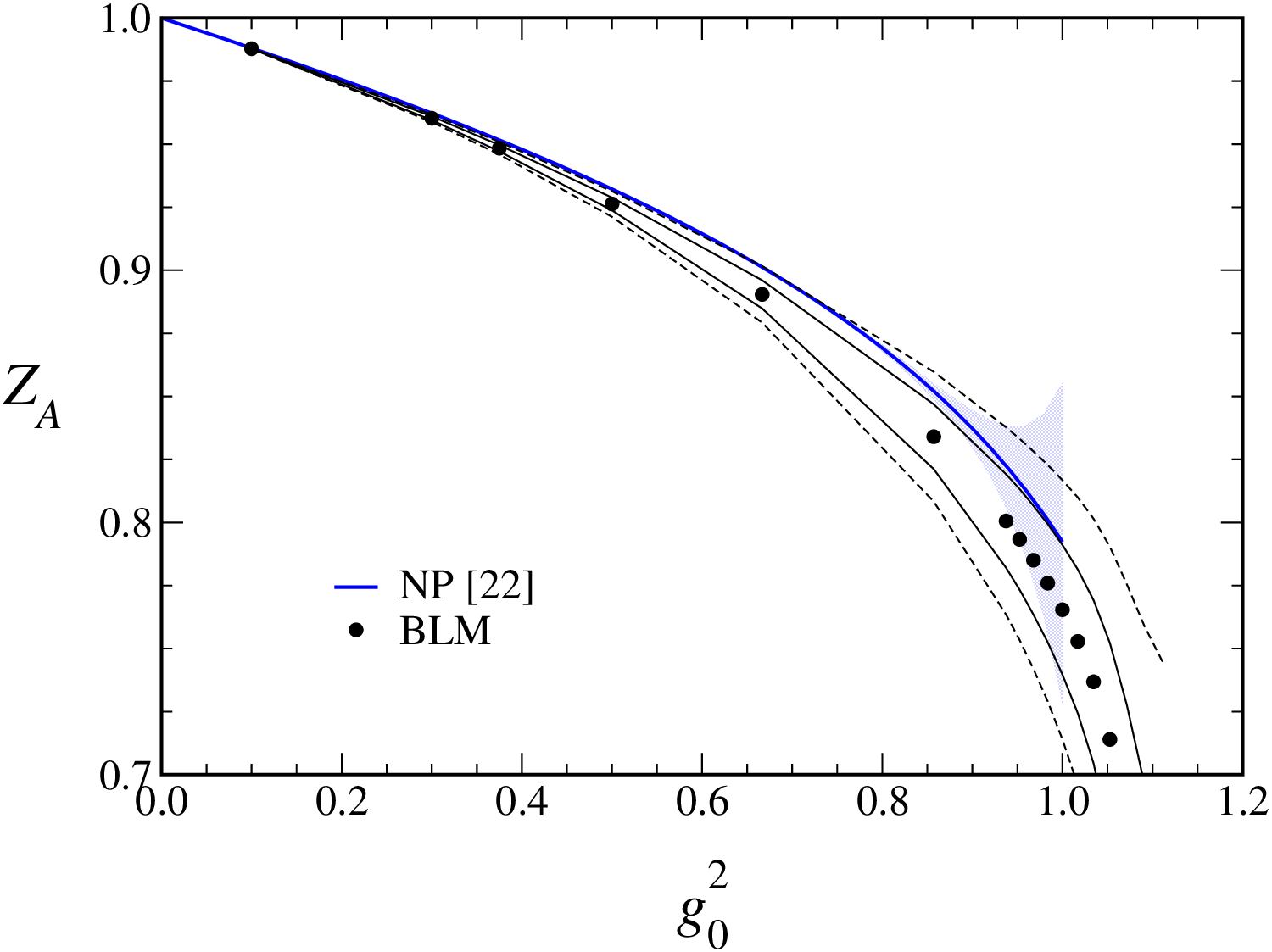} \hfill
	(d)\hspace*{-1.2em}
	\includegraphics[width=0.48\textwidth]{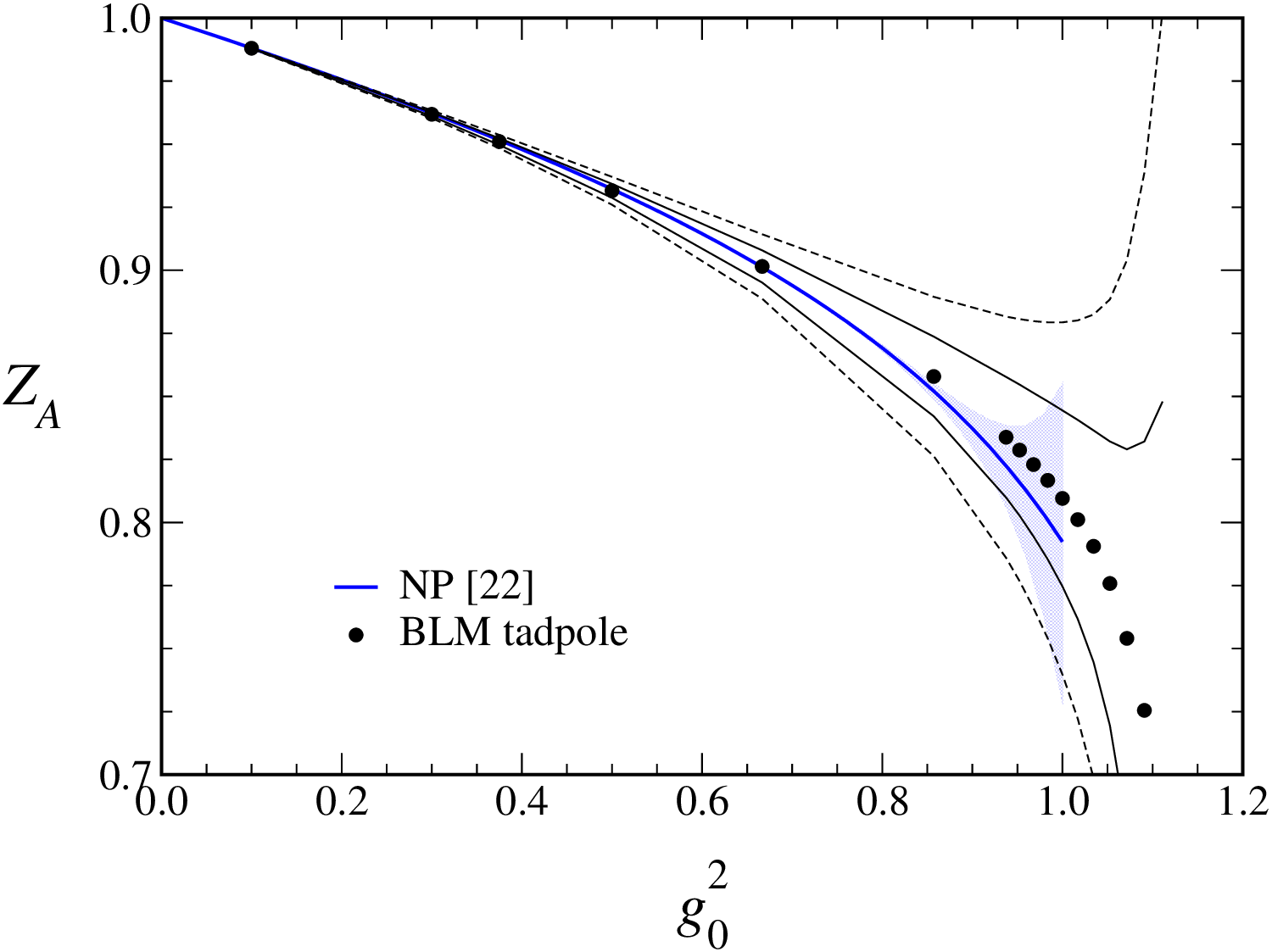}
	\caption[fig:ZJ]{$Z_V$ and $Z_A$ vs.\ $g_0^2$.
	Heavy lines show the non-perturbative results,
	Eqs.~(\ref{eq:ZVPade}) and~(\ref{eq:ZAPade}),
	and shading possible corrections of order $\pm(\Lambda a)^2$.
	Circles show BLM perturbation theory, with 
	thin and dashed lines to indicate a two-loop term
	$\pm\alpha_V^2$ or $\pm2\alpha_V^2$.
	(a) and (c) no tadpole improvement,
	$Z_J^{\text{BLM}}=1+g_V^2(q^*_{Z_J})Z_J^{[1]}$;
	(b) and (d) with tadpole improvement,
	$Z_J^{\text{BLM}}=u_0[1+g_V^2(q^*_{\tilde{Z}_J})\tilde{Z}_J^{[1]}]$.}
	\label{fig:ZJ}
\end{figure*}

For the normalization factors $Z_V$ and $Z_A$, BLM perturbation theory
and the non-perturbative methods agree well, within 3--4\%.
The difference between the two non-perturbative values of~$Z_V$ exceeds
the reported errors, but is easily explained by power correction of
order~$(\Lambda a)^2$.
For the tadpole-free ratio $Z_A/Z_V$ and for the
tadpole-improved quantities $u_0\tilde{Z}_J$,
BLM perturbation theory lies very close to the non-perturbative range.
These impressions are strengthened by Fig.~\ref{fig:ZJ}, which shows
$Z_V$ and $Z_A$ as functions of~$g_0^2$.
Circles show BLM perturbation theory, and the thin solid (dashed)
lines indicate how two-loop contributions of $\pm\alpha_V^2$
($\pm2\alpha_V^2$) could modify the result.
We show the result with and without tadpole improvement in
Figs.~\ref{fig:ZJ}(b,d) and~(a,c), respectively.
For the non-perturbative method, a heavy (blue) line shows the Pad\'e
approximants~\cite{Luscher:1996ax}
\begin{eqnarray}
	Z_V & = & \frac{1-0.7663g_0^2+0.0488g_0^4}{1-0.6369g_0^2},
	\label{eq:ZVPade} \\
	Z_A & = & \frac{1-0.8496g_0^2+0.0610g_0^4}{1-0.7332g_0^2},
	\label{eq:ZAPade}
\end{eqnarray}
which deviate from the underlying calculations negligibly for
$g_0^2\le1$.
The shaded bands behind the Pad\'e curves show a power-correction of
$\pm(\Lambda a)^2$, with~$\Lambda\sim500~\textrm{GeV}$.
The finite-volume result also suffers from power corrections of order
$(a/L)^2$.
They are estimated to be small by comparing calculations on
lattices with $a/L=1/8$ and~$1/12$~\cite{Luscher:1996ax}.
Also, they are parametrically smaller, because Ref.~\cite{Luscher:1996ax}
holds $L\Lambda\sim2$ for all~$g_0^2$.

\begin{figure*}[btp]
	(a)\hspace*{-1.2em} \includegraphics[width=0.48\textwidth]{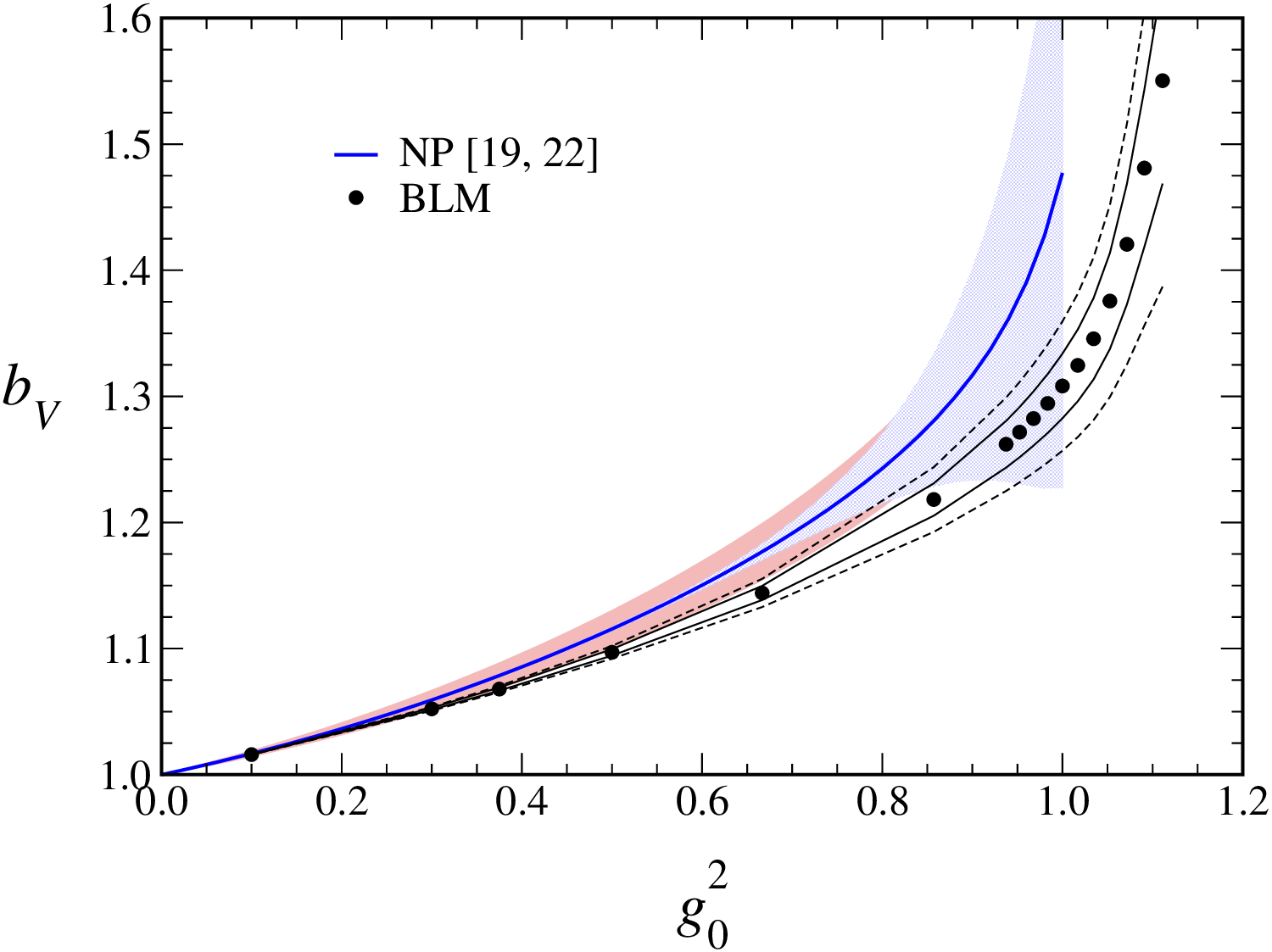} \hfill
	(b)\hspace*{-1.2em} \includegraphics[width=0.48\textwidth]{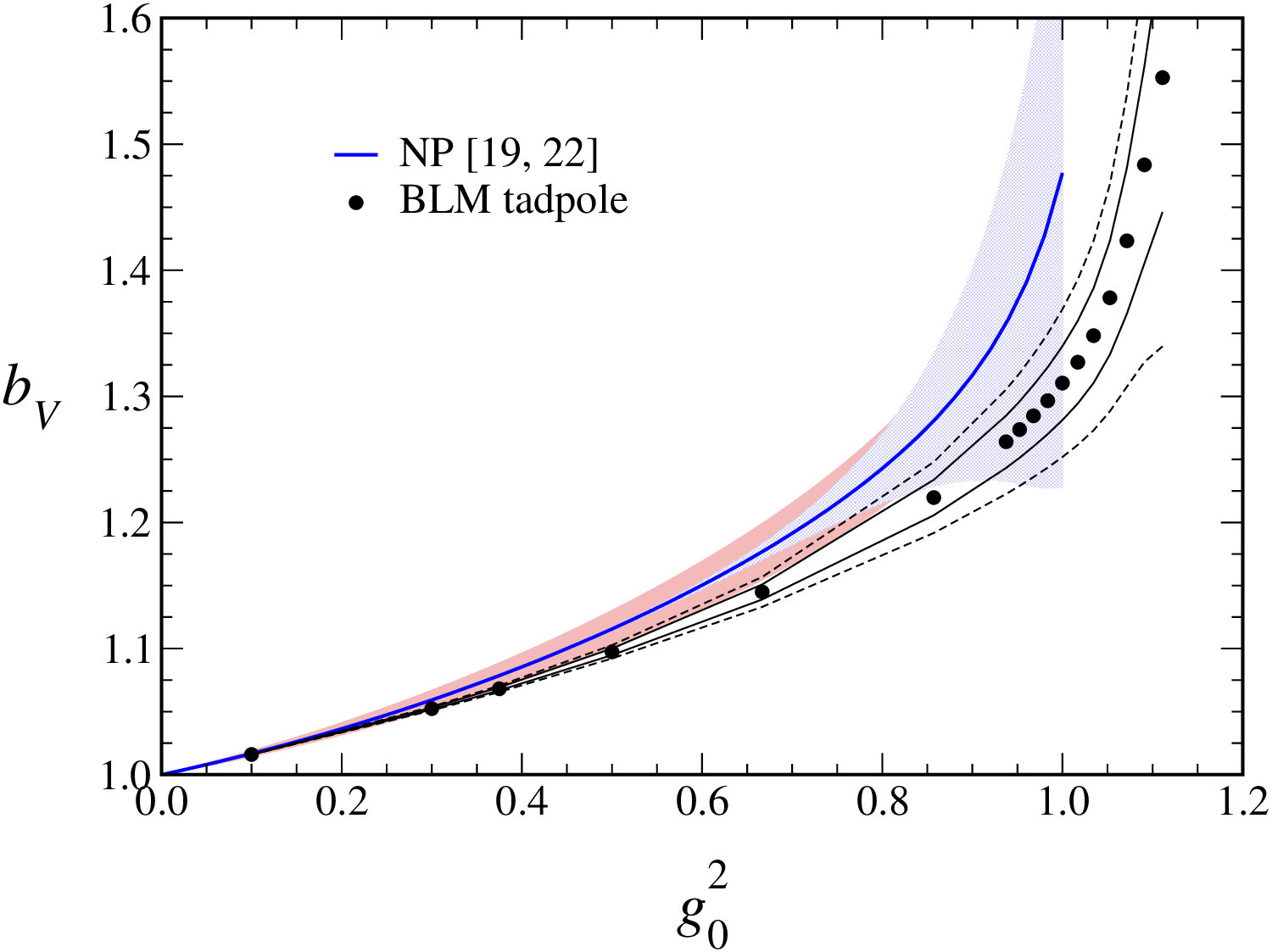}
	\caption[fig:bV]{$b_V$ vs.\ $g_0^2$.
	(a) no tadpole improvement,
	$b_V^{\text{BLM}}=1+g_V^2(q^*_{b_V})b_V^{[1]}$;
	(b) with tadpole improvement,
	$b_V^{\text{BLM}}=[1+g_V^2(q^*_{\tilde{b}_V})\tilde{b}_V^{[1]}]/u_0$.
	Light grey (light blue) shading indicates power corrections to $b_V$
	of order $\pm\Lambda a$;
	darker grey (pink) shading power corrections to $b_V-1$ of order
	$\pm a/L$.}
	\label{fig:bV}
\end{figure*}

Next let us turn to the $O(ma)$ corrections to the normalization
factors, $b_V$ and $b_A$.
There is only one calculation of $b_A$~\cite{Bhattacharya:2001pn},
so let us concentrate first on~$b_V$.
The two non-perturbative results for $b_V$ agree perfectly with each
other (see the Tables), but they deviate significantly from
one-loop BLM perturbation theory.
Some insight can be gleaned from Fig.~\ref{fig:bV}, which shows $b_V$
as a function of~$g_0^2$.
The non-perturbative method is represented with the Pad\'e
approximant~\cite{Sint:1997jx}
\begin{equation}
	b_V = \frac{1-0.7613g_0^2+0.0012g_0^4-0.1136g_0^6}{1-0.9145g_0^2},
	\label{eq:bVPade}
\end{equation}
with light (blue) shading for a power correction~$\pm\Lambda a$.
In finite volume there is also a power correction to~$b_V$ of order~$a/L$;
by construction it applies to $b_V-1$~\cite{Luscher:1996ax},
but now $L$ with $a$ varies such that $a/L=1/8$ for all~$g_0^2$.
We model this effect as $(b_V-1)(1\pm\frac{1}{8})$, shown in the darker
(pink) shading in Fig.~\ref{fig:bV}.
Judging from its size and shape, the deviation seen in Fig.~\ref{fig:bV}
looks less like a two-loop effect than a combination of power
corrections of order $a/L$ and~$\Lambda a$.
(Similar conclusions are reached in Ref.~\cite{Bhattacharya:2001pn}.)
There is almost no difference whether one applies tadpole improvement
to~$b_V$ or not, once the BLM prescription is applied.
These two approximations truncate higher orders of the perturbation
series differently, substantiating the idea that the discrepancy is
a power correction.

The non-perturbative calculation of $b_A$ agrees with one-loop BLM
perturbation theory.
Note, however, that Ref.~\cite{Bhattacharya:2001pn} obtains $b_V$
and~$b_V-b_A$ directly, and then $b_A=b_V-(b_V-b_A)$.
The agreement between BLM perturbation theory and the non-perturbative
results for $b_V$ and $b_V-b_A$ is not good, so the agreement for $b_A$
may be an accident.
Since the coefficient $b_V^{[1]}-b_A^{[1]}$ in Eq.~(\ref{eq:bV-bA})
is remarkably small, the two-loop contribution could be as large as
the one-loop term.
Furthermore, inspection of Fig.~14 in Ref.~\cite{Bhattacharya:2001pn}
suggests that a fit to the three smallest masses would yield a smaller
value of $b_V-b_A$.
We consider the comparison of $b_A$ and $b_V-b_A$ to be unsettled
pending a two-loop calculation and a more robust non-perturbative
calculation.

In any case, the mild disagreement on $b_V$ and $b_V-b_A$ is not of
much practical importance.
For the sake of argument, suppose $ma<0.1$, which holds for the light
quarks for which the currents were designed.
Then power corrections in $b_J$, at fixed $a$, lead to an uncertainty
in a decay constant or a form factor of only a few per cent.
After a continuum limit extrapolation, these uncertainties will not be
important.

\begin{figure*}[btp]
	(a)\hspace*{-1.2em} \includegraphics[width=0.48\textwidth]{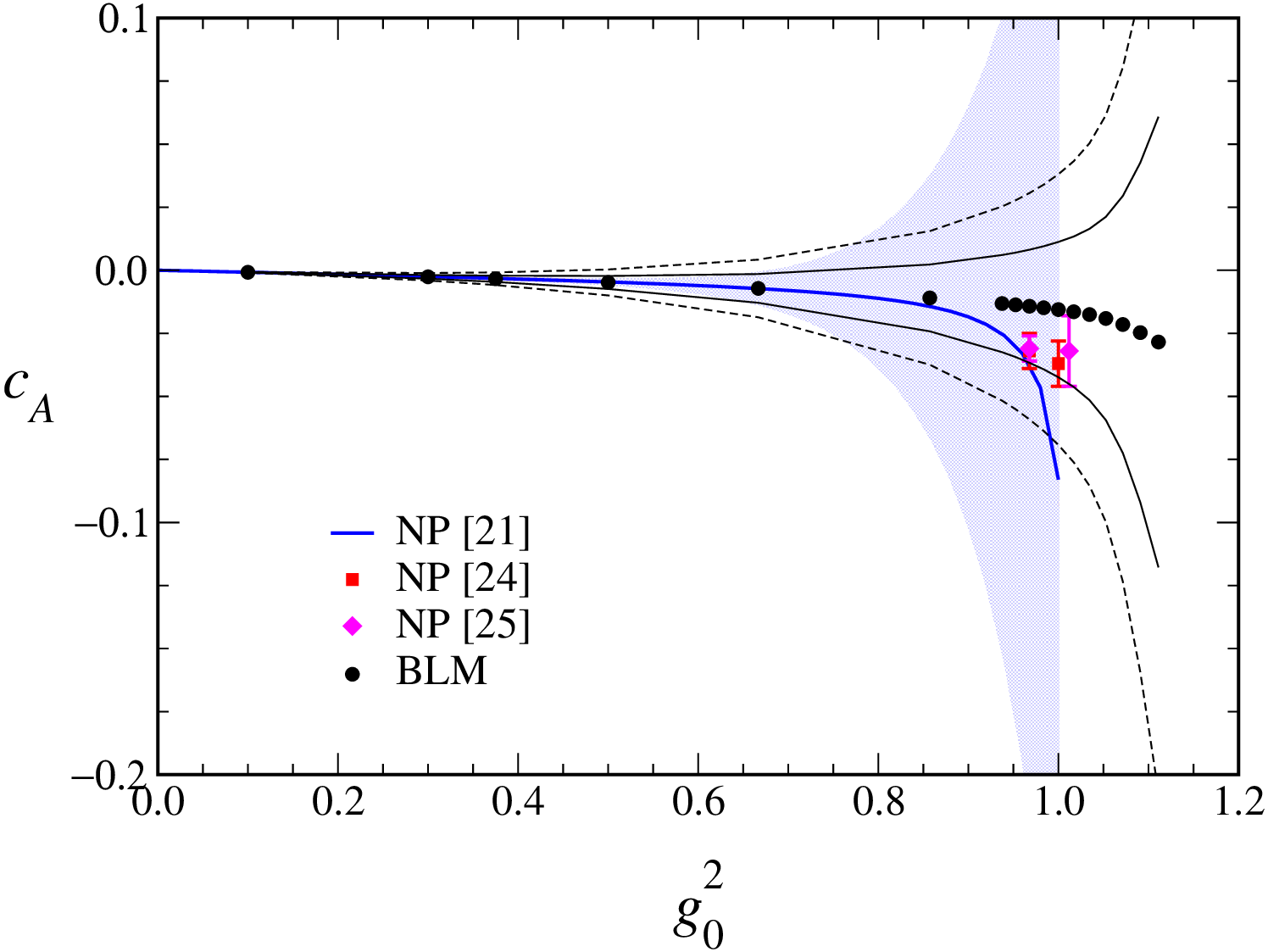}
	(b)\hspace*{-1.2em} \includegraphics[width=0.48\textwidth]{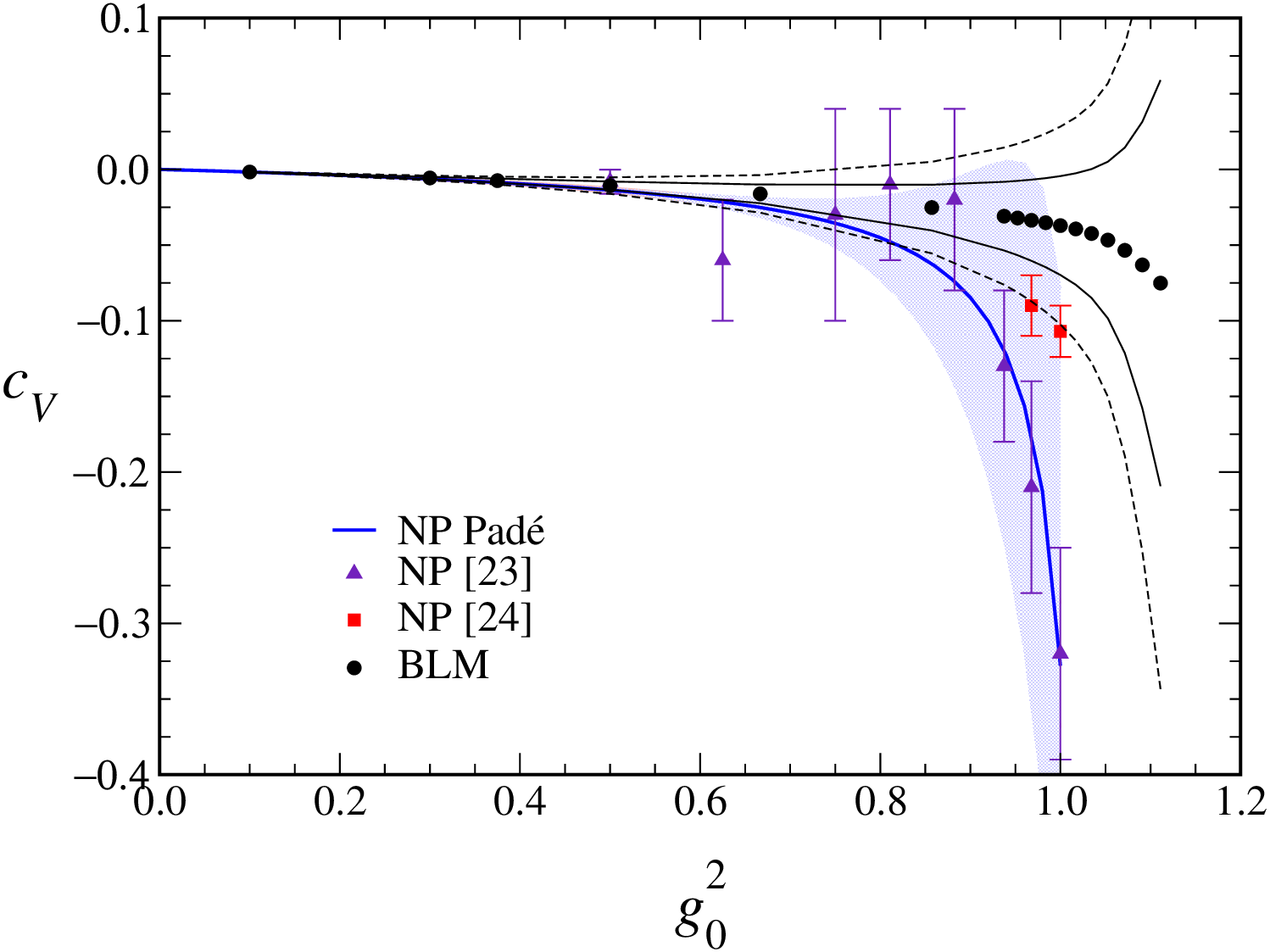}
	\caption[fig:cJ]{$c_A$ and $c_V$ vs.\ $g_0^2$.
	Shading shows power corrections of order~$\pm\Lambda a$ to 
	(a) Eq.~(\ref{eq:cAPade}),
	(b) Eq.~(\ref{eq:cVPade}).
	Points with error bars are from
	(a) Refs.~\cite{Bhattacharya:2001pn,Collins:2001mm},
	(b) Refs.~\cite{Guagnelli:1997db,Bhattacharya:2001pn}.}
	\label{fig:cJ}
\end{figure*}

Now let us turn to the coefficients~$c_J$ of the improvement terms in
Eq.~(\ref{eq:Vlat}) and~(\ref{eq:Alat}).
At the tabulated values of $\beta$, the non-perturbative and BLM
calculations of $c_A$ do not agree at all.
At $\beta=6.0$ (Table~\ref{tab:bK60}) the two non-perturbative
calculations also do not agree with each other.
Figure~\ref{fig:cJ}(a) shows $c_A$ as a function of~$g_0^2$,
using the Pad\'e approximant~\cite{Luscher:1996ck}
\begin{equation}
	c_A = -0.00756g_0^2\frac{1-0.748g_0^2}{1-0.977g_0^2}
	\label{eq:cAPade}
\end{equation}
to represent the non-perturbative calculations.
The disagreement between BLM perturbation theory and
Eq.~(\ref{eq:cAPade}) sets in for $g_0^2>0.9$.
There are two reasons to suspect that the discrepancy stems
from a power correction of order $\Lambda a$ to the results of
Ref.~\cite{Luscher:1996ck}.
First, Fig.~\ref{fig:cJ}(a) shows that it has the shape and size of such
a power correction.
Second, the extracted value of $c_A$ depends on the lattice
derivative used to define the current~\cite{Collins:2001mm}.
Note~\cite{Bhattacharya:2001pn} that errors in $c_A$ propagate to
$c_V$, because in the Ward identities $c_A$ is multiplied by large
hadronic matrix elements such as $am^2_K/m_s\sim a\times2.5~\textrm{GeV}$.
This enhancement also explains why Eq.~(\ref{eq:cAPade}) leads to worse
scaling in $f_\pi$~\cite{Collins:2001mm}.
Figure~\ref{fig:cJ}(a) also includes the non-perturbative results of
Refs.~\cite{Bhattacharya:2001pn,Collins:2001mm}.
The difference between those points and BLM perturbation theory could
be a modest two-loop effect or a small power correction.

For~$c_V$, the two non-perturbative results agree neither with each
other, nor with BLM perturbation theory.
The {\sl Alpha} Collaboration has only a preliminary
calculation~\cite{Guagnelli:1997db}.
We have taken the liberty of extracting results from Fig.~3 of
Ref.~\cite{Guagnelli:1997db} and fitting them to a Pad\'e formula.
The leading behavior is fixed to Eq.~(\ref{eq:cV1}), and we obtain
\begin{equation}
	c_V = -0.01633g_0^2\frac{1-0.257g_0^2}{1-0.963g_0^2}.
	\label{eq:cVPade}
\end{equation}
Figure~\ref{fig:cJ}(b) plots Eq.~(\ref{eq:cVPade}), the underlying
points~\cite{Guagnelli:1997db}, the non-perturbative results from
hadronic correlation functions~\cite{Bhattacharya:2001pn}, and BLM
perturbation theory.
As usual we show possible power corrections to Eq.~(\ref{eq:cVPade}) of
order~$\pm\Lambda a$, as well as the size of typical two-loop effects.
At small $g_0^2$, there is good agreement with (BLM) perturbation
theory, but once $g_0^2>0.9$, there is a sharp turnover.
It is probably a power correction, possibly exacerbated by power
corrections to $c_A$ as modeled by Eq.~(\ref{eq:cAPade}).
With hadronic correlation functions~\cite{Bhattacharya:2001pn}
the non-perturbative value of $c_V$ is half or a third as large.
It is not clear at present whether the discrepancy between
Ref.~\cite{Bhattacharya:2001pn} and BLM perturbation theory is a power
correction to the former or a sizable two-loop correction to the latter.

We should also mention that BLM perturbation theory works better than
several forms of mean-field perturbation theory (let alone bare
perturbation theory).
In Table~\ref{tab:alpha} we list several choices for $\alpha_s$:
\begin{eqnarray}
	\alpha_0          = g_0^2/4\pi, \\
	\tilde{\alpha}_0  = \alpha_0/u_0, 
\end{eqnarray}
as well as $\alpha_{1\times1}$ [Eq.~(\ref{eq:1x1})] and
$\alpha_{\overline{\rm MS}}(q^*_{\overline{\rm MS}})$
[Eq.~(\ref{eq:MS})].
With only one-loop expansions available, the mean-field choices
$\tilde{\alpha}_0$ and $\alpha_{1\times1}$ give smaller corrections,
and one-loop perturbation theory falls short even when power corrections
are negligible.
The consistency of BLM-$V$ perturbation theory for $Z_V$, $Z_A$, and
$Z_A/Z_V$ indicates that the BLM prescription does indeed re-sum an
important class of higher-order contributions.
On the other hand, the coupling
$\alpha_{\overline{\rm MS}}(q^*_{\overline{\rm MS}})$
seems, empirically, to work less well.
In continuum perturbative QCD, it usually does not matter whether one
adopts $\alpha_V(q^*_V)$,
$\alpha_{\overline{\rm MS}}(q^*_{\overline{\rm MS}})$ or some other
renormalized coupling (at the BLM scale), once two-loop effects are
included.
It would not be surprising for the same to hold for short-distance
quantities in lattice gauge theory, such as improvement coefficients.
\begin{table}[t!]
	\centering
	\caption[tab:alpha]{Expansion parameters for perturbation theory.}
	\begin{ruledtabular}
	\begin{tabular}{llllll}
		$\beta$ & $\alpha_0$ & $\tilde{\alpha}_0$ & $\alpha_{1\times1}$
		& $\alpha_V(q^*_{Z_V})$ &
		$\alpha_{\overline{\rm MS}}(q^*_{Z_V})$ \\
		\hline
		6.0 & 0.0796 & 0.1340 & 0.1245 & 0.1602 & 0.1784 \\
		6.2 & 0.0770 & 0.1255 & 0.1166 & 0.1468 & 0.1619 \\
		6.4 & 0.0746 & 0.1183 & 0.1101 & 0.1362 & 0.1491 \\
		7.0 & 0.0682 & 0.1016 & 0.0951 & 0.1134 & 0.1222 \\
		9.0 & 0.0531 & 0.0702 & 0.0667 & 0.0748 & 0.0786 \\
	\end{tabular}
	\end{ruledtabular}
	\label{tab:alpha}
\end{table}

\section{Conclusions}
\label{sec:conclusions}

In this paper we have compared non-perturbative calculations of several
improvement coefficients to perturbation theory with the BLM
prescription.
Previously this could not be done, because the ``BLM numerators'' in
Eqs.~(\ref{eq:starZV})--(\ref{eq:starcA}) were not available.
We find that, for the scale-independent quantities considered here,
the integration of the $\log k^2$-weighted integrals is numerically
straightforward.

BLM perturbation theory for the current normalization factors~$Z_J$
agrees very well with non-perturbative calculations of the same
quantities.
Here the leading power correction is only of order $(\Lambda a)^2$,
and the small deviations can probably be removed with a two-loop
calculation.
Note that generalizations of the BLM method for higher-order
perturbation theory have been considered in continuum
perturbative QCD~\cite{Brodsky:1994eh} and in lattice gauge
theory~\cite{Hornbostel:2000ey}.

For the improvement coefficients $b_J$ and $c_J$,
the leading power corrections are of order $\Lambda a$
(and in the Schr\"odinger functional also of order $a/L=1/8$),
while some of the one-loop coefficients are small.
It is consequently difficult to diagnose the discrepancies.
By noting the size and dependence on $g_0^2$
of the differences, we concur with the authors of
Refs.~\cite{Bhattacharya:2001pn,Collins:2001mm}, namely, that power
corrections contaminate the non-perturbative results.
In particular, it seems unlikely that higher orders in
perturbative series could explain all discrepancies between
one-loop BLM perturbation theory and the results from
Refs.~\cite{Luscher:1996ck,Luscher:1996ax,Guagnelli:1997db}.


\begin{thebibliography}{99}
%
\bibitem{Symanzik:1979ph}
K. Symanzik,
in \emph{Recent Developments in Gauge Theories},
edited by G. 't~Hooft \emph{et al}.\ (Plenum, New York, 1980).
%
\bibitem{Symanzik:1983dc}
K. Symanzik,
in \emph{Mathematical Problems in Theoretical Physics},
edited by R. Schrader \emph{et al}.\ (Springer, New York, 1982);
Nucl.\ Phys.\ B {\bf 226}, 187, 205 (1983).
%
\bibitem{Wilson:1975hf}
K. G. Wilson,
in {\em New Phenomena in Subnuclear Physics},
edited by A. Zichichi (Plenum, New York, 1977).
%
\bibitem{Sheikholeslami:1985ij}
B.~Sheikholeslami and R.~Wohlert,
Nucl.\ Phys.\  B {\bf 259}, 572 (1985).
%
\bibitem{Kronfeld:2002pi}
For a review, see A.~S.~Kronfeld,
in \emph{At the Frontier of Particle Physics: Handbook of~QCD},
Vol.~4, edited by M.~Shifman (World Scientific, Singapore, 2002)
[arXiv:hep-lat/0205021].
%
\bibitem{Jansen:1996ck}
K.~Jansen {\it et al.},
Phys.\ Lett.\ B {\bf 372}, 275 (1996)
[arXiv:hep-lat/9512009].
%
\bibitem{Luscher:1996sc}
M.~L\"uscher, S.~Sint, R.~Sommer and P.~Weisz,
Nucl.\ Phys.\ B {\bf 478}, 365 (1996)
[arXiv:hep-lat/9605038].
%
\bibitem{Martinelli:1997zc}
G.~Martinelli, G.~C.~Rossi, C.~T.~Sachrajda, S.~R.~Sharpe, M.~Talevi and
M.~Testa,
Phys.\ Lett.\ B {\bf 411}, 141 (1997)
[arXiv:hep-lat/9705018].
%
\bibitem{deDivitiis:1997ka}
G.~M.~de Divitiis and R.~Petronzio,
Phys.\ Lett.\ B {\bf 419}, 311 (1998)
[arXiv:hep-lat/9710071].
%
\bibitem{Bhattacharya:1999uq}
T.~Bhattacharya, S.~Chandrasekharan, \hfill R.~Gupta, W.~J.~Lee and
S.~R.~Sharpe,
Phys.\ Lett.\ B {\bf 461}, 79 (1999)
[arXiv:hep-lat/9904011].
%
\bibitem{Davier:1997kw}
M.~Davier and A.~H\"ocker,
Phys.\ Lett.\ B {\bf 419}, 419 (1998)
[arXiv:hep-ph/9711308].
%
\bibitem{Lepage1991:zt}
G.~P.~Lepage and P.~B.~Mackenzie,
Nucl.\ Phys.\ B Proc.\ Suppl.\  {\bf 20}, 173 (1991).
%
\bibitem{Brodsky:1983gc}
S.~J.~Brodsky, G.~P.~Lepage and P.~B.~Mackenzie,
Phys.\ Rev.\  D {\bf 28}, 228 (1983).
%
\bibitem{Lepage:1993xa}
G.~P.~Lepage and P.~B.~Mackenzie,
Phys.\ Rev.\  D {\bf 48}, 2250 (1993)
[arXiv:hep-lat/9209022].
%
\bibitem{Harada:2001fi}
J.~Harada, S.~Hashimoto, K.~I.~Ishikawa, A.~S.~Kronfeld, T.~Onogi
and N.~Yamada,
Phys.\ Rev.\ D {\bf 65}, 094513 (2002)
[arXiv:hep-lat/0112044].
%
\bibitem{Harada:2001fj}
J.~Harada, S.~Hashimoto, A.~S.~Kronfeld and T.~Onogi,
Phys.\ Rev.\ D {\bf 65}, 094514 (2002)
[arXiv:hep-lat/0112045].
%
\bibitem{Gabrielli:1990us}
E.~Gabrielli, G.~Martinelli, C.~Pittori, G.~Heatlie and C.~T.~Sachrajda,
Nucl.\ Phys.\ B {\bf 362}, 475 (1991).
%
\bibitem{Capitani:2001xi}
S.~Capitani, M.~G\"ockeler, R.~Horsley, H.~Perlt, P.~E.~Rakow,
G.~Schierholz and A.~Schiller,
Nucl.\ Phys.\ B {\bf 593}, 183 (2001)
[arXiv:hep-lat/0007004].
%
\bibitem{Sint:1997jx}
S.~Sint and P.~Weisz,
Nucl.\ Phys.\ B {\bf 502}, 251 (1997)
[arXiv:hep-lat/9704001].
%
\bibitem{Bernard:1998sx}
C.~W.~Bernard, M.~Golterman and C.~McNeile,
Phys.\ Rev.\ D {\bf 59}, 074506 (1999)
[arXiv:hep-lat/9808032].
%
\bibitem{Luscher:1996ck}
M.~L\"uscher, S.~Sint, R.~Sommer, P.~Weisz and U.~Wolff,
Nucl.\ Phys.\ B {\bf 491}, 323 (1997)
[arXiv:hep-lat/9609035].
%
\bibitem{Luscher:1996ax}
M.~L\"uscher, S.~Sint, R.~Sommer and H.~Wittig,
Nucl.\ Phys.\ B {\bf 491}, 344 (1997)
[arXiv:hep-lat/9611015].
%
\bibitem{Guagnelli:1997db}
M.~Guagnelli and R.~Sommer,
Nucl.\ Phys.\ B Proc.\ Suppl.\  {\bf 63}, 886 (1998)
[arXiv:hep-lat/9709088].
%
\bibitem{Bhattacharya:2001pn}
T.~Bhattacharya, R.~Gupta, W.~Lee and S.~Sharpe,
Phys.\ Rev.\ D {\bf 63}, 074505 (2001)
[arXiv:hep-lat/0009038].
%
\bibitem{Collins:2001mm}
S.~Collins, C.~T.~H.~Davies, G.~P.~Lepage and J.~Shi\-ge\-mitsu,
arXiv:hep-lat/0110159.
%
\bibitem{Brodsky:1994eh}
S.~J.~Brodsky and H.~J.~Lu,
Phys.\ Rev.\ D {\bf 51}, 3652 (1995)
[arXiv:hep-ph/9405218].
%
\bibitem{Hornbostel:2000ey}
K.~Hornbostel, G.~P.~Lepage and C.~Morningstar,
Nucl.\ Phys.\ B Proc.\ Suppl.\  {\bf 94}, 579 (2001)
[arXiv:hep-lat/0011049].
%
\end{thebibliography}
\end{document}